\documentclass[journal]{IEEEtran}

\usepackage{mathpazo}
\usepackage{amsmath}
\usepackage{graphicx}
\usepackage{multirow}
\usepackage{cite}
\usepackage{graphicx}
\usepackage{footnote}
\usepackage{bbding}
\usepackage{pifont}
\usepackage{wasysym}
\usepackage{amssymb}
\usepackage{multirow}
\usepackage[table,xcdraw]{xcolor}
\usepackage{diagbox}
\usepackage[flushleft]{threeparttable}
\makesavenoteenv{tabular}
\usepackage{tabu}
\usepackage{longtable}
\usepackage{epstopdf}
\usepackage{color}
\usepackage{enumitem}
\usepackage{booktabs}
\usepackage{subfigure}
\setlist{parsep=0pt,listparindent=\parindent}
\usepackage{caption}
\usepackage[latin1]{inputenc}

\ifCLASSINFOpdf
\else
\fi

\usepackage{ragged2e}
\usepackage{verbatim}
\begin{document}
%
\title{Multiscale Evolutionary Perturbation Attack on Community Detection}
%
%
%
%

\author{Jinyin~Chen, Yixian~Chen, Lihong~Chen, Minghao~Zhao, and~Qi~Xuan,~\IEEEmembership{Member,~IEEE}%
\thanks{ This work was supported in part by the National Natural Science Foundation of China under Grant 61502423, Grant 61572439 and Grant 62072406, in part by the Zhejiang Provincial Natural Science Foundation of China under Grant LR19F030001 and Grant LY19F020025, in part by the Zhejiang Science and Technology Plan Project under Grant LGF18F030009, and in part by the Key Technologies, System and Application of Cyberspace Big Search, Major Project of Zhejiang Lab under Grant 2019DH0ZX01, the Major Special Funding for "Scien"ce and Technology Innovation 2025" in Ningbo under Grant No. 2018B10063.
 \emph{(Corresponding author: Qi Xuan.)}}
\IEEEcompsocitemizethanks{
\IEEEcompsocthanksitem J. Chen, Y. Chen, L. Chen, and Q. Xuan are with the Institute of Cyberspace Security, College of Information Engineering, Zhejiang University
of Technology, Hangzhou 310023, China. E-mail: \{chenjinyin, 2111803168, 2111803032, xuanqi\}@zjut.edu.cn.\protect\\
M. Zhao is with the Fuxi AI Lab, NetEase Inc, Hangzhou 310052, China. E-mail: zhaominghao@corp.netease.com.\protect\\}
}

%
%

\maketitle
%



\IEEEtitleabstractindextext{%
\begin{abstract}
\justifying
Community detection, aiming to group nodes based on their connections, plays an important role in network analysis, since communities, treated as meta-nodes, allow us to create a large-scale map of a network to simplify its analysis. However, for privacy reasons, we may want to prevent communities from being discovered in certain cases, leading to the topics on community deception. In this paper, we formalize this community detection attack problem in three scales, including global attack (macroscale), target community attack (mesoscale) and target node attack (microscale). We treat this as an optimization problem and further propose a novel Evolutionary Perturbation Attack (EPA) method, where we generate adversarial networks to realize the community detection attack. Numerical experiments validate that our EPA can successfully attack network community algorithms in all three scales, i.e., hide target nodes or communities and further disturb the community structure of the whole network by only changing a small fraction of links. By comparison, our EPA behaves better than a number of baseline attack methods on six synthetic networks and three real-world networks. More interestingly, although our EPA is based on the louvain algorithm, it is also effective on attacking other community detection algorithms, validating its good transferability.
\justifying
\end{abstract}

}

\maketitle

\IEEEdisplaynontitleabstractindextext

%
\IEEEpeerreviewmaketitle

\ifCLASSOPTIONcompsoc
\IEEEraisesectionheading{\section{Introduction}\label{sec:introduction}}
\else
\section{Introduction}
\label{sec:introduction}
\fi

%
%
%
%
\IEEEPARstart{C}{onsisting} of many nodes and links, network captures certain relationship in real world and is often used as a mathematical representation for a variety of complex systems, such as social systems~\cite{chen2017interplay,Ding2017TeamGen,fu2018link,xuan2018social}, transportation systems~\cite{an2018network,bao2017planning} and supply chain systems~\cite{xuan2011framework,he2018time}, etc. Many real-world networks can be divided into communities, with the nodes within the same communities connected densely, while those across different communities connected sparsely~\cite{malliaros2013clustering}. The revealed community structure can not only present the close relationship among the nodes inside a certain community, but may also indicate that these nodes tend to share common properties or play similar roles in the respective fields~\cite{kamdar2017phlegra}. Analyzing community structure in a network thus can help to better understand the interactions inter- and intra- close groups of nodes in the network, which is the exact reason why new community detection algorithms are continuously proposed and widely used in a large number of areas.

Community detection algorithms are designed to divide the network into partitions of dense regions which correspond to strong related entities. Evaluating the effectiveness of these algorithms has been a controversial problem until Newman proposed the modularity $Q$~\cite{newman2004finding,newman2016community}.
The proposal of modularity makes the problem of non-overlapping community detection unprecedented developed. A bunch of modularity optimization algorithms were subsequently proposed, some of which are based on splitting or aggregation~\cite{newman2004fast,girvan2002community,clauset2005finding,schuetz2008multistep}, while many others use optimization algorithms to maximizes the modular $Q$, such as annealing~\cite{guimera2005functional}, Particle Swarm Optimization (PSO)~\cite{zhou2016neighborhood}, external optimization~\cite{folino2014evolutionary} and spectral optimization~\cite{fortunato2010community}. These algorithms translate community detection into an optimization problem and try to find the optimal community division by maximizing certain fitness. In addition, there are also community detection algorithms based on information theory~\cite{rosvall2008maps,pons2005computing,alamgir2010multi,fu2013threshold} and label propagation~\cite{raghavan2007near,barber2009detecting,liu2010advanced}. The former  believes that data flow can be compressed with regular code, such as random walk model. The main idea is that the probability of wandering to the nodes in the same group is much greater than those in different groups; while the latter updates each node label to its most frequent neighbor label through iteration, which was widely used due to its simplicity and high efficiency. 
\begin{table*}[!t]
\centering
\caption{Comparison with the existent approaches.}
\begin{tabular}{@{}ccccc@{}}
\toprule
     Approach   & Scale of Attack &  Way to Change Network & Budget Required in the beginning  \\
     \midrule
  Nagaraja~\cite{nagaraja2010impact} & Mesoscale &  Addition & Yes \\
  Waniek et al.~\cite{waniek2018hiding} & Mesoscale&	Rewiring & Yes  \\
   Fionda et al.~\cite{fionda2018community} & Mesoscale &  Rewiring & Yes \\
   Chen et al.~\cite{chen2019ga} & Macroscale &  Rewiring & Yes \\
   Li et al.~\cite{li2020adversarial} & Macroscale &  Rewiring & Yes \\
   Our EPA  & Macroscale \& Mesoscale \& Microscale &  Rewiring \& Addition & No\\
  \bottomrule
\end{tabular}
\label{table:compare}
\end{table*}

However, on one hand, due to privacy concerns, people may not want their social information, such as being a member of certain communities, to be discovered by algorithms; 
on the other hand, a lot of graph-based models, e.g., graph-based recommenders, need to integrate community detection to improve their efficiency, especially for those relatively large systems. The normal operation of such systems thus may rely on the robustness of community detection algorithm. These bring up a problem: how to attack community detection algorithms or defend against such attacks? Since different links play different roles in keeping the community structure of a network, the community detection algorithms could be significantly disturbed by only changing a small fraction of links. In this paper, we focus on the attack part, and seek the maximum community change by rewiring minimal number of links, which can help to judge which links are most vulnerable and thus also provides insights for the defense part in order to keep the community structure.

There are some attack strategies to disturb network algorithms~\cite{zugner2018adversarial,bojchevski2019adversarial,chen2018fast,yu2019target}, but few studies related to community detection attack. Nagaraja~\cite{nagaraja2010impact} first introduced a community hiding problem, where they added links based on centrality values. Waniek et al.~\cite{waniek2018hiding} proposed a heuristic rewiring method to hide a community, by deleting the links within the same communities while adding the links across different communities. Because the links are random selected, the method is of relatively low effectiveness. And they also propose the concept of individual hiding, as a way to avoid an influential node being highlighted by three centrality measures (i.e., degree, closeness, betweenness). Fionda et al.~\cite{fionda2018community} proposed a deception score to evaluate the effect of community deception, which takes the accuracy and recall of community detection into account, as well as the reachability of target community. The essence of this score is that the target community is expected to be divided into more new communities and each new community contains less percentage of target nodes. However, this may lead to the following two problems: first, it encourages more communities which might make the attack less concealed; second, when the number of communities is set to be constant, it can't guide the nodes from the target community to the optimal community to achieve better attack capacity. Quite recently, Chen et al.~\cite{chen2019ga} proposed Q-Attack based on Genetic Algorithm (GA) and modularity, for the first time, to disturb the overall modular structure of a network. Li et al.~\cite{li2020adversarial} generated adversarial networks by attacking the graph neural networks (GNNs) based surrogate model to hide a set of nodes from community detection.

Generally, we think there are following three types of community detection attacks of different scales.
\begin{itemize}
\item \textbf{Global attack (macroscale):} This attack is to achieve maximum community changes of the whole network by rewiring minimal number of links.
\item \textbf{Target community attack (mesoscale):} This attack is also known as community deception, which is to hide one specific community by rewiring minimal number of links that are connected to at least one node in the community.
\item \textbf{Target node attack (microscale):} This attack is to make target node belong to different communities by rewiring the minimal number of links around it. Note that our target node attack is different from the individual hiding proposed by Waniek et al.~\cite{waniek2018hiding}, since the latter has nothing to do with community detection.
\end{itemize}

In this paper, we formalize the community detection attack problem and present an Evolutionary Perturbation Attack (EPA) algorithm to attack community detection algorithms in three scales,  i.e., macroscale (network), mesoscale (community) and microscale (node). The main differences between our method and the previous attack algorithms are summarized in TABLE~\ref{table:compare}. Moreover, we use a series of metrics to evaluate the attack results obtained by different attack methods on several real datasets. In particular, we make the following main contributions.
\begin{itemize}
\item To the best of our knowledge, this is the first time to formalize the problem of community detection attack into three scales, i.e., macroscale (network), mesoscale (community) and microscale (node).
\item We propose a novel EPA algorithm to attack community detection, which is capable of generating approximate optimal adversarial network with the minimal number of rewired links to launch all the three scales of attacks. We compare our EPA with other baseline attack methods on several synthetic networks and real-world networks, and find that EPA behaves the best in most cases, achieving the state-of-the-art results.
\item We use GA to solve the optimization problem in EPA algorithm, and meanwhile introduce network structural properties including betweenness and the shortest path lengths between pairwise nodes into the mutation process to accelerate the convergence rate, making it faster to obtain the approximate optimal solution.
\item We find that our EPA has outstanding transferability, i.e., the adversarial network generated by EPA against the louvain algorithm (LOU) can also be used to successfully fool other community detection algorithms.
\end{itemize}

The rest of paper is organized as follows. In Sec.~\ref{sec:EPA}, we formalize the problem of community detection attack and introduce our EPA method in details. Then, we compare the attack effects by utilizing EPA and other attack methods on a number of synthetic networks and real-world networks in Sec.~\ref{sec:Experiments}, where we further use a variety of community detection algorithms to verify the transferability of EPA. Finally, we conclude the paper and highlight future research directions in Sec.~\ref{sec:Conclusions}.

\section{Community Detection Attack\label{sec:EPA}}
\subsection{Problem Formulation\label{PF}}
The main symbols used in this paper are listed in TABLE~\ref{table:variable} for convenience. First, we state and formalize the problem of community detection attack. Given a network $G$=$(V,E)$ as an undirected graph including a set of nodes $V$ and a set of links $E$, suppose it can be divided into communities $C$=$\{C_1,C_2,\cdots,C_p\}$, with $C_i\subseteq V$ and $C_i\cap C_j$=$\emptyset$, for all $i$,$j\in \{1,2,\cdots,p\}$ and $i\neq j$. Then, we propose the concept of community detection attack, which is launched by an adversarial network to fool community detection methods. We study this problem in three scales: global attack (macroscale), target community attacks (mesoscale) and target node attack (microscale). Under these definitions, some other definitions and problems can be defined as follows.

\begin{table}[!t]
 \newcommand{\tabincell}[2]{\begin{tabular}{@{}#1@{}}#2\end{tabular}}
 \centering
 \caption{The notation definition of main symbols.}
 \setlength{\tabcolsep}{0mm}{
 \begin{tabular}{@{}lr@{}}
 \toprule
      \textbf{Symbol}   & \textbf{Definition}   \\
      \midrule
      $G$&The original network\\
      $V/E/A$&The nodes/links/adjacency matrix of original network\\
      $\hat{G}/\bar{G}$&The perturbation/adversarial network\\
      $\hat{E}/\bar{E}$&The links of perturbation/adversarial network\\
      $\hat{A}/\bar{A}$&The adjacency matrix of perturbation/adversarial network\\
      $E^{+}/E^{-}$&The set of link additions/deletions\\
      $\phi_{g}/\phi_{t}$&The evaluation function of global/target community attack\\
      $\delta/\mu$&The evaluation function of community/network modification\\
      $\epsilon$&The predefined threshold of target node attack\\
   $n$&The number of nodes in network\\
   $m/m_t$&The number of links in network/target community\\
   $\beta$& The budget which is the number of rewired links\\
   $\theta$&The maximum number of rewired links\\
   $C/\bar{C}$&The communities before/after the attack\\
   $M$&The confusion matrix whose element is $m_{i,j}$\\
   $M_{i.}/M_{.i}$&The number of nodes in new/original community $C_{i}$\\
   $m_{i,j}/\tilde{m}_{i,j}$&\tabincell{r}{The number of community $C_{i}$'s nodes which originally \\belong/not belong to the target community $C_{j}$}\\
   $E_{M_r}/E_{M_c}$& The entropy of new/target communities\\
   $N_a/N_b$& The number of communities in the control/test group\\
   $O_i$&The $i$-th chromosome of the population\\
   $(.)|O_i$&The value under the attack represented by $O_i$ \\
   \bottomrule
 \end{tabular}}
 \label{table:variable}
 \end{table}

\begin{center}
\begin{figure*}[!t]
\includegraphics[width=\linewidth]{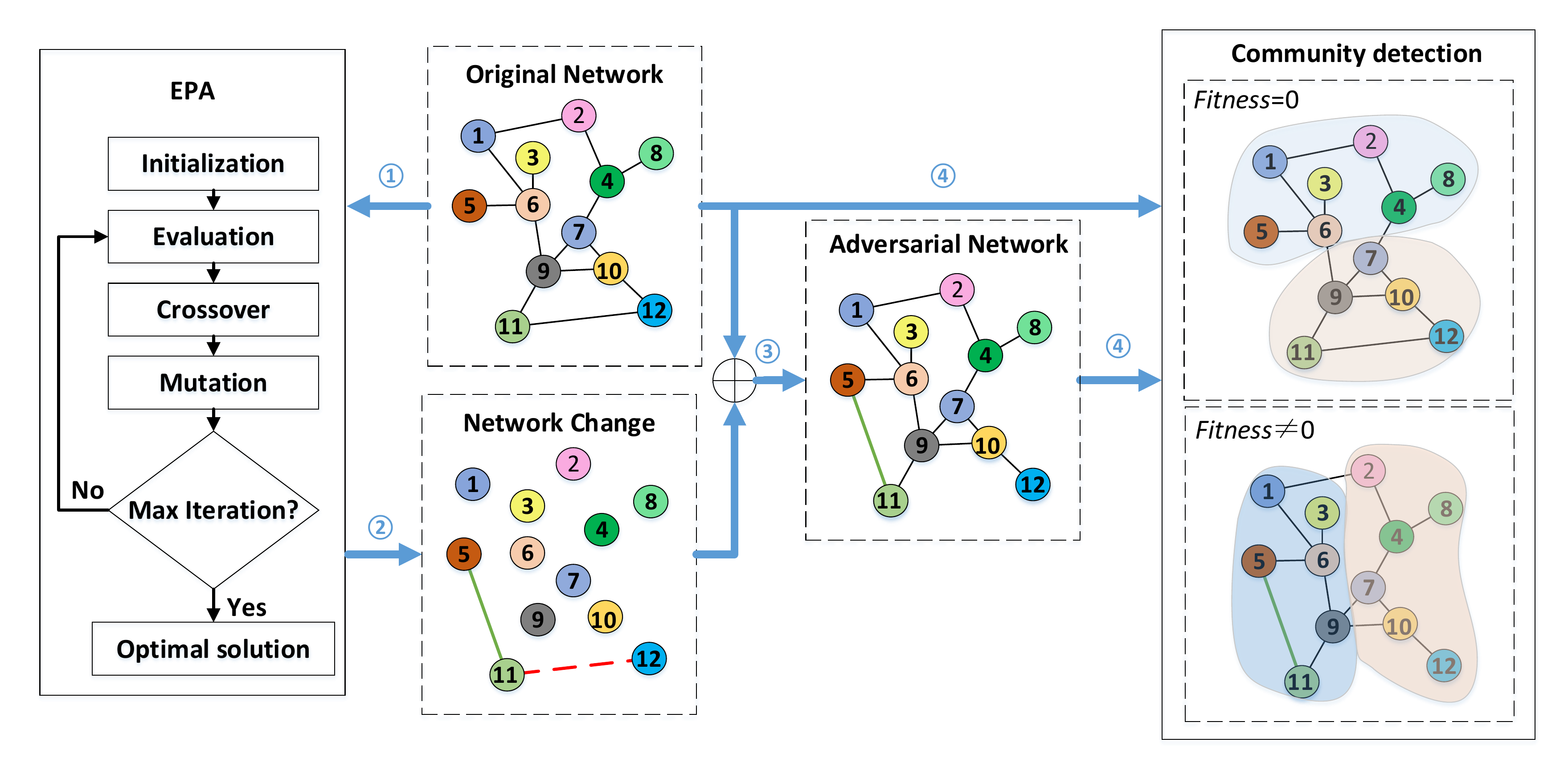}
\centering
\caption{The framework of EPA, which consists of four steps. First, an original network is input to EPA. Second, GA is used to generate the approximate optimal network change, denoted by the sets of added and deleted links. Third, an adversarial network is established by changing the links in the original network. Finally, various community detection algorithms are performed on the adversarial network to validate the attack effect of EPA.}
\label{fig:frame}
\end{figure*}
\end{center}

\newtheorem{Problem}{Problem}
\newtheorem{Definition}{Definition}
\begin{Definition}[\textbf{Adversarial network}]
Denoting the original network $G$=$(V,E)$ as the target, the adversarial attack selects some key links to construct an adversarial perturbation network $\hat{G}$=$(V,\hat{E})$, and $\hat{A}$ is the adjacency matrix whose element $\hat{A}_{u,v}\in[-1,0,1]$ means the modified strategy on the adjacency matrix of original network $A$.  Adversarial network $\bar{G}$=$(V,\bar{E})$, generated by adding adversarial perturbation on the original network, has the adjacency matrix $\bar{A}$ defined as
\begin{equation}
\bar{A}=A+\hat{A}.
\label{adversarial_net}
\end{equation}
\end{Definition}
\begin{Problem}[\textbf{Global Attack}]
Given a network $G$=$(V,E)$, generate an adversarial network $\bar{G}$=$(V,\bar{E})$ via attack strategy to make community detection method fail with budget $\beta$. The function $\phi_{g}$ is used to evaluate the attack gain of global attack, and $\bar{G}$ is divided into communities $\bar{C}$=$\{\bar{C_1},\bar{C_2},\cdots,\bar{C_q}\}$ by using the same community detection algorithm. Furthermore, $E^{+}$ (resp., $E^{-}$) denotes a set of link additions (resp., deletions) and the global attack is realized by solving the optimization problem:
\begin{equation}
	\mathop{\arg\max}{\{ \phi_{g}(C,\bar{C},E,\bar{E},E^{+},E^{-},\beta)\}}\label{op1},
\end{equation}
where $\beta = \left|E^{+}\right| =\left|E^{-}\right|$, meaning that we make the total number of links remains the same, and $\bar{E}=(E \cup E^{+}) \backslash E^{-}$. Since the links are rewired globally, we have
\begin{equation}
	E^{+} \subseteq \{(u,v) : u \in V \wedge v \in V,(u,v) \notin E\},
\end{equation}
\begin{equation}
	E^{-} \subseteq \{(u,v) : u \in V \wedge v \in V,(u,v) \in E\}.
\end{equation}
\end{Problem}

\begin{Problem}[\textbf{Target Community Attack}]
Given a target community $C_i\subseteq C$, which is obtained by community detection method $\varphi$ on the original network $G$. After target community attack, $C_i$ cannot be detected by $\varphi$, i.e., in the adversarial network $\bar{G}$, the nodes in $C_i$ belong to a set of new communities, denoted by $\dot{C}$=$\{\dot{C_1},\dot{C_2},\cdots,\dot{C_r}\} \subseteq \bar{C}$, with $\dot{C_j} \cap C_i \neq \emptyset, \forall j \in [1,r]$. Given the function $\phi_{t}$ which is used to evaluate the attack gain of target community attacks, and it is realized by solving the optimization problem:
\begin{equation}
	\mathop{\arg\max}{\{ \phi_{t}(C_i,\dot{C},E,\bar{E},E^{+},E^{-},\beta)\}}\label{op2}.
\end{equation}
Here, we still consider the rewiring process and thus have $\beta = \left|E^{+}\right|=\left|E^{-}\right|$, $\bar{E}=(E \cup E^{+}) \backslash E^{-}$. In order to hide community $C_{i}$, it would be best to disconnect the links inside the community while add connections between the nodes in $C_i$ and those in other communities, thus we have
\begin{equation}
	E^{+} \subseteq \{(u,v) : u \in C_i \oplus v \in C_i,(u,v) \notin E\},
\end{equation}
\begin{equation}
	E^{-} \subseteq \{(u,v) : u \in C_i \wedge v \in C_i,(u,v) \in E\}.
\end{equation}
\end{Problem}

\begin{Problem}[\textbf{Target Node Attack}]
Given a target node $t$, suppose it belongs to community $C_{i}$ in the original network $G$, while it belongs to $\bar{C_j}$ in the adversarial network $\bar{G}$, then the target node attack is realized by solving the optimization problem:
\begin{eqnarray}
&\mathop{\arg\max}{\{\delta \times \mu(E,\bar{E},E^{+},E^{-},\beta)\}},
\label{node-1}\\
&\delta =  \left\{
    \begin{array}{cc}
    1 & \frac{\left|\bar{C}_j \cap C_i\right|}{\left|\bar{C}_j\right|} \textless \epsilon \\
    0 & else
    \end{array},\right.
\label{node-2}
\end{eqnarray}
where $\delta$ and $\mu$ are the functions to measure the amount of community and network change, respectively. $\epsilon \in [0,1]$ is a predefined threshold, based on which we determine whether communities $C_i$ and $\bar{C_j}$ are close to each other or not. We think the attack is successful only when these two communities are relatively different from each other. Similarly , we still have $\beta = \left|E^{+}\right|=\left|E^{-}\right|$ and $\bar{E}=(E \cup E^{+}) \backslash E^{-}$. To make the attack more effective, at this time, we rewire links around the target node $t$, thus we have
\begin{equation}
	E^{+} \subseteq \{(u,t) : u \in V ,(u,t) \notin E\},
\end{equation}
\begin{equation}
	E^{-} \subseteq \{(u,t) : u \in V ,(u,t) \in E\}.
\end{equation}
\end{Problem}

Note that the attack gain consists of two parts: the budget $\beta$, defined as the number of rewired links, and the attack effect. As we can see, the overall attack effect generally increases as the budget $\beta$ grows, as a result, it is our focus to improve the attack gain with only limited budget $\beta$.

These indicate that community detection attack can always be represented as an optimization problem, which could be basically solved by evolutionary computing methods, such as Genetic Algorithm (GA). In this paper, we propose a GA based Evolutionary Perturbation Attack (EPA) for community detection. In order to make the GA more suitable for finding appropriate attack strategy, we use rewired link ID as genes on the chromosome instead of binary coding, which can effectively reduce the storage space of the population. Considering that the number of rewired links is also a variable here, we adopt a strategy called \emph{unequal crossover} to make the length of chromosome changeable during the crossover process. Moreover, a novel search mechanism based on network structural properties, including betweenness and the shortest path lengths between pairwise nodes, is introduced in the mutation process to accelerate the convergence of GA, making it faster to obtain the approximate optimal solution.

In particular, our EPA is established in four stages, including initialization, evaluation, crossover and mutation, which will be introduced one by one in the following. The flowchart of EPA is shown in Fig.~\ref{fig:frame}.

\subsection{Initialization\label{Ini}}
\begin{figure}[!t]
\includegraphics[width=\linewidth]{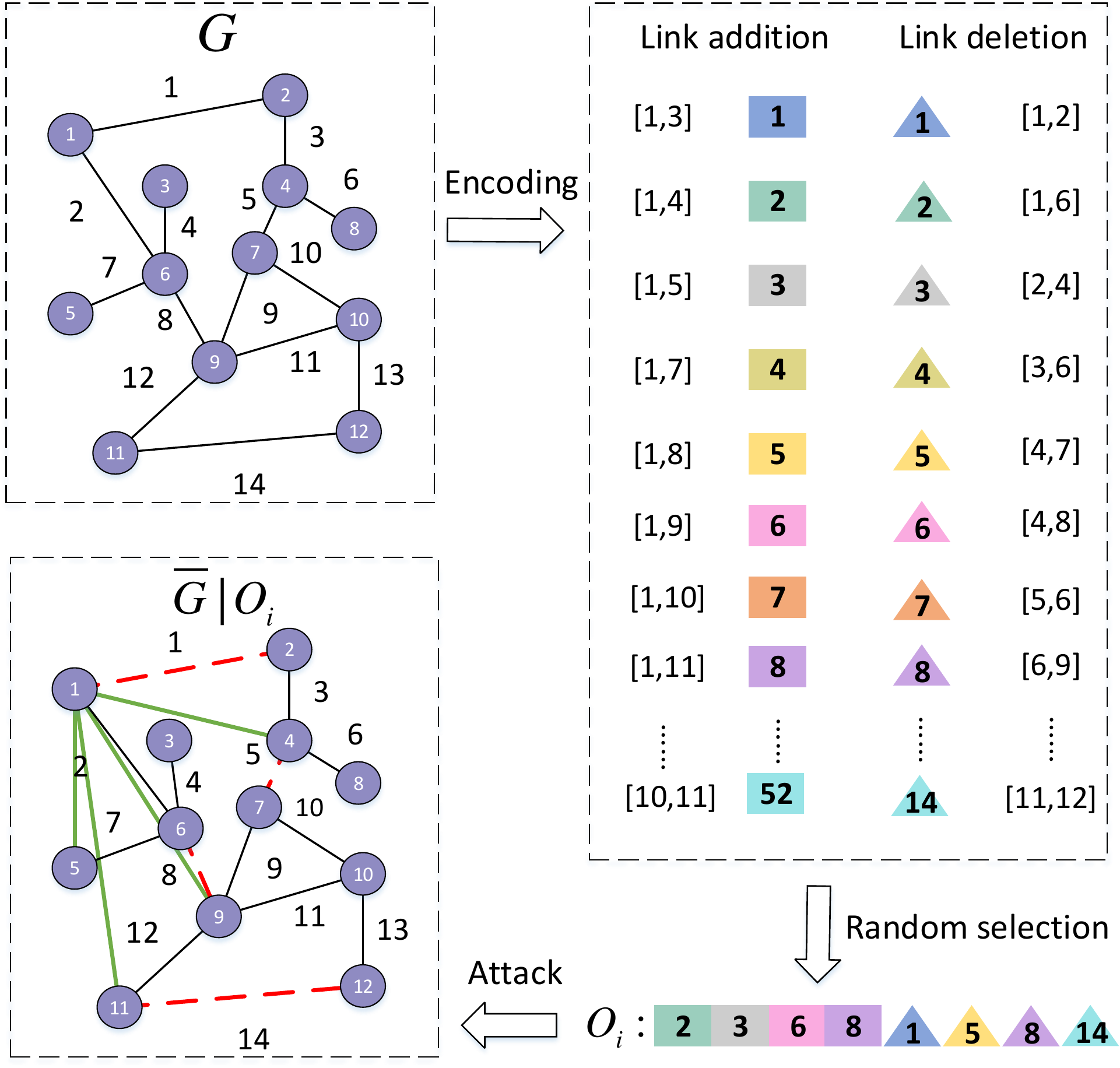}
\centering
\caption{An illustration of chromosome encoding and initialization. $\bar{G}|O_i$ is the adversarial network after the attack represented by chromosome $O_i$. The original network $G$ contains 14 links and 52 disconnected pairs of nodes that have potential to be connected in the attack. The red dashed lines and the green solid lines represent the links that are deleted and added in the attack, respectively. In the left, the rectangles represent the link addition genes while the triangles denote the link deletion genes. Then, we initialize chromosome $O_i$ by randomly selecting $\beta=4$ link addition and deletion genes, respectively.}
\label{fig:Initialization}
\end{figure}

First, we directly use the ID of rewired link as the gene of chromosome to facilitate the evolving. Specifically, we create the indexes for existent links and nonexistent links and treat them as link deletion and addition genes, respectively. To make the attack more concealed, we set the number of deleted links equal to that of added links, so that the total number of links in the network keeps constant. Denote the threshold of the number of rewired links by $\theta$ and the chromosome by $O_{i}$ with the budget $\beta \in [1, \theta] $. We thus have
\begin{eqnarray}
O_{i} &=& \{A_i, B_i\} \nonumber \\
    &=&\{a_i^1,a_i^2,\cdots,a_i^\beta,b_i^1,b_i^2,\cdots,b_i^\beta\},
\label{equ:1}
\end{eqnarray}
where $O_i$, with the index $i\in[1,P]$, represents the $i$-th chromosome in population and $P$ is the population size of GA. $A_i$ and $B_i$ represent the sets of link addition and deletion links, respectively, in chromosome $O_i$, with the element $a_i^k$ (or $b_i^k$) being the index of a node pair which are disconnected (or connected). The adjacency matrix of adversarial perturbation network $\hat{A}$, represented by chromosome $O_i$, is thus calculated by
\begin{eqnarray}
\hat{A}_{uv}=\left\{
    \begin{array}{cc}
    1 & Index(u,v) \in A_i \\
    -1& Index(u,v) \in B_i\\
    0 & \mbox{otherwise}
    \end{array},\right.
\label{AF-2}
\end{eqnarray}
where $Index(u,v)$ is the function to obtain the index of a node pair $(u,v)$. An illustration to explain how chromosomes are coded and initialized is shown in Fig.~\ref{fig:Initialization}.

Note that, to make the target node attack more effective, instead of initializing randomly,
we first select a community $C_t$ randomly, then we preferentially delete the links between the target node and those not belonging to the community $C_t$, while establish links between the target node and those in the community $C_t$. If all nodes in the community $C_t$ are connected to the target node, then we randomly select another community and repeat the above steps.

\subsection{Evaluation\label{Eva}}
\subsubsection{Fitness Function\label{Fitness}}
Each chromosome corresponds to an attack strategy. After encoding the attack strategies, we then need to evaluate the attack effect of each strategy using some fitness function. Note that the entropy is maximized when each new community consists of nodes uniformly from many different \emph{real} communities. Since the uniform distribution emphasized by entropy makes the overall accuracy and recall as low as possible, we thus think it's an appropriate metric to evaluate the attack effect. For the global attack or target community attack, their fitness functions follow the same general form, represented by
\begin{eqnarray}
\phi_{i} = \Psi(d'|O_i) \times X(C,\bar{C}|O_i),
\label{overview-1}
\end{eqnarray}
which consists of two parts: the attenuation function $\Psi\in[0,1]$ and the attack effect evaluation function $X$. $\Psi$ is a monotonic decreasing function of $d'|O_i$, with $d'$ being the normalization of $d$ which represents the degree change after the attack represented by chromosome $O_i$. $X$ is the function to evaluate the attack effect based on entropy, and the terms  $C$ and $\bar{C}|O_i$ denote the community detection results before and after the attack, respectively. For the target node attack, however, since the attack effect is binary, i.e., either success or failure, the fitness function only contains the first part, i.e., the change of target node degree.

\subsubsection{Attenuation Function\label{AF}}
In order to perform the attack with limited budget $\beta$, we use the exponential decay function as our attenuation function, described as follows:
\begin{eqnarray}
\Psi(d'|O_i) = \exp (c \times d')|O_i,
\label{AF-1}
\end{eqnarray}
where $d'=d/m$ ($m$ is the number of links in the whole network) for global attack while $d'=d/m_t$ ($m_t$ is the number of links in the target community) for target community attack; and $c$ is a constant that controls the decay speed.
For a network of $n$ nodes, the degrees of all nodes before and after the attack are denoted by $D $=$ \{d_1,d_2,\cdots,d_n \}$ and $\bar{D}$=$\{\bar{d}_1,\bar{d}_2,\cdots,\bar{d}_n \}$, respectively. Then, the distance $d$ between them is calculated by
\begin{eqnarray}
d = \frac{1}{4}\sum^{n}_{i=1}\left|d_i-\bar{d}_i\right|,
\label{AF-3}
\end{eqnarray}
where $d_i$ and $\bar{d}_i$ are the degrees of node $i$ before and after the attack, respectively.

We think that, for the same budget $\beta$, the fewer nodes change their degrees, the smaller the network structure varies. We thus use distance $d$ instead of budget $\beta$ as the input of attenuation function. In the case of $\beta=1$, the degrees of at most four nodes will change, therefore the value of $d$ defined by Eq.~(\ref{AF-3}) tends to be 1, equal to the number of rewired links. When the rewiring involves the same node, the distance $d$ will be smaller than budget $\beta$. Our EPA method are thus more likely to rewire links involving fewer nodes, making it more concealed. 


\subsubsection{Attack Effect\label{AEE}}
Given a confusion matrix $M$ with each element $m_{ij}$ representing the number of the shared nodes between the original community $C_{i}$ and the new community $\bar{C}_j$, we define the function $X$ to evaluate the attack effect:
\begin{eqnarray}
X(C,\bar{C}|O_i) = (\mathop{E_{M_r}'}+\mathop{E_{M_c}'})|O_i,
\label{AEE-1}
\end{eqnarray}
where $\mathop{E_{M_r}'}$ and $\mathop{E_{M_c}'}$ are the normalized entropy of $M_r$ and $M_c$, and thus $X$ must be in the range of $[0,2]$. Suppose $E_{M_r}$ and $E_{M_c}$ are the entropy of new communities and target communities, which are obtained by considering the row vectors and column vectors of $M$, respectively. Suppose the matrix $M$ dimension is $N_a\times N_b$, with $N_a$ and $N_b$ being the numbers of communities in control group (i.e., real community number) and test group (i.e., community number detected by certain community detection algorithm after the attack), respectively,
the maximum values of $E_{M_r}$ (resp., $E_{M_c}$) $E_{M_c}$ are obtained when the values of each row (resp., column) are equal. In this case, the calculation of $E_{M_r}$ and $E_{M_c}$ is simplified to
\begin{eqnarray}
\mathop{\max}E_{M_r} = -\sum^{N_a}_{i=1}\frac{M_{i.}}{n}(\frac{1}{N_b}\log_{2}\frac{1}{N_b} \times N_b)=\log_{2}N_b,
\label{M-1}\\
\mathop{\max}E_{M_c} = -\sum^{N_b}_{i=1}\frac{M_{.i}}{n}(\frac{1}{N_a}\log_{2}\frac{1}{N_a} \times N_a)=\log_{2}N_a,
\label{M-2}
\end{eqnarray}
where $n$ is the number of nodes in network. $M_{i.}$ and $M_{.i}$ represent the numbers of nodes in new and original community $C_{i}$, respectively.

For global attack, 
the entropy for new communities $E_{M_r}$ and that for target communities $E_{M_c}$ are calculated by
\begin{eqnarray}
E_{M_r} = -\sum^{N_a}_{i=1}\sum^{N_b}_{j=1}\frac{M_{i.}}{n}(\frac{m_{i,j}}{M_{i.}}\log_{2}\frac{m_{i,j}}{M_{i.}}),
\label{global-1}\\
E_{M_c} = -\sum^{N_b}_{i=1}\sum^{N_a}_{j=1}\frac{M_{.i}}{n}(\frac{m_{j,i}}{M_{.i}}\log_{2}\frac{m_{j,i}}{M_{.i}}),
\label{global-2}
\end{eqnarray}
where $m_{ij}$ is the number of nodes in new community $C_{i}$ that originally belong to community $C_{j}$.

Suppose all the communities keep exactly the same after the attack, we have $E_{M_r}=E_{M_c}=0$. With Eq.~(\ref{M-1}) we can conclude that $E_{M_r}$ and $E_{M_c}$ must be in the range of $[0,\log_{2}{N_b}]$ and $[0,\log_{2}{N_a}]$, respectively. Therefore, based on Eq.~(\ref{overview-1}) and Eq.~(\ref{AEE-1}), the fitness of $O_i$ for global attack is defined as
\begin{eqnarray}
\phi_{i} = \Psi(d'|O_i) \times (\frac{E_{M_r}}{\log_{2}{N_b}}+\frac{E_{M_c}}{\log_{2}{N_a}})|O_i.
\label{global-5}
\end{eqnarray}

For target community attack, here we limit that any added or deleted link must be connected to at least one node in the target community, which greatly reduces the searching space of solutions. In this case, $M$ is an $N_a\times2$ matrix with element $m_{ij}$ (resp., $\tilde{m}_{i,j}$) being the number of nodes in community $C_{i}$ that originally belong (resp., don't belong) to target community $C_{j}$. For target community $C_{j}$, $E_{M_r}$ and $E_{M_c}$ are defined as
\begin{alignat}{2}
E_{M_r} &= -\sum^{N_a}_{i=1}\frac{M_{i.}}{n}(\frac{m_{i,j}}{M_{i.}}\log_{2}\frac{m_{i,j}}{M_{i.}}+\frac{\tilde{m}_{i,j}}{M_{i.}}\log_{2}\frac{\tilde{m}_{i,j}}{M_{i.}}),
\label{tc-1}\\
E_{M_c} &= -\sum^{N_a}_{i=1}\frac{m_{i,j}}{M_{.i}}\log_{2}\frac{m_{i,j}}{M_{.i}}.
\label{tc-2}
\end{alignat}
Based on the above definitions, we always have $\tilde{m}_{i,j}$=$M_{i.}-m_{i,j}$. The remaining variables are defined the same as those in Eq.~(\ref{global-1}) and Eq.~(\ref{global-2}).

\begin{figure*}[!t]
\includegraphics[width=\linewidth]{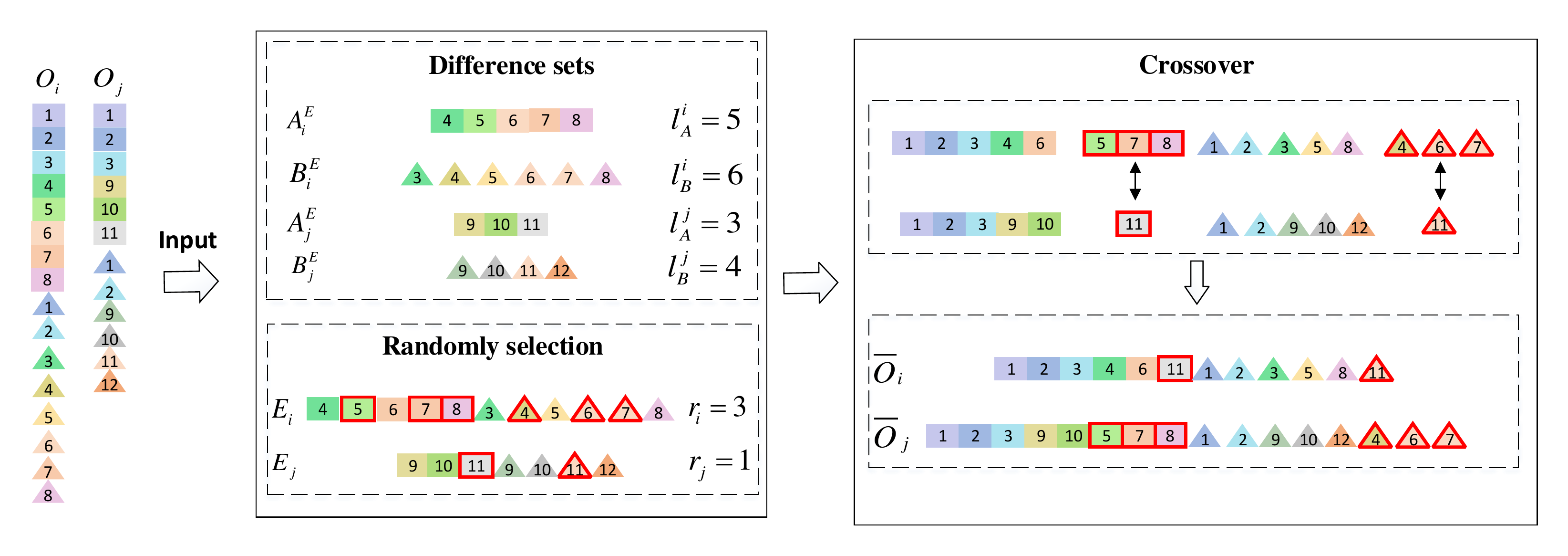}
\centering
\caption{An example of non-equal crossover operation. $O_i$ and $O_j$ (resp., $\bar{O}_i$ and $\bar{O}_j$) are the chromosomes before (resp., after) the crossover. $l_A^i$ and $l_B^i$ (resp. $l_A^j$ and $l_B^j$) are the numbers of exchangeable addition and deletion genes, respectively, on chromosome $O_i$ (resp. $O_j$). Furthermore, $r_i$ and $r_j$ are the numbers of crossed genes in $O_i$ and $O_j$, respectively.}
\label{fig:crossover}
\end{figure*}

Similarly, with Eq.~(\ref{M-1}) we can conclude that $E_{M_r}$ and $E_{M_c}$ must be in the range of $[0,1]$ and $[0,\log_{2}{N_a}]$, respectively. Therefore, the fitness of $O_i$ for target community attack is defined as
\begin{eqnarray}
\phi_{i} = \Psi(d') \times(E_{M_r}+\frac{E_{M_c}}{\log_{2}N_a}).
\label{tc-5}
\end{eqnarray}
It should be noted that our fitness function based on entropy is somewhat similar to the deception score $H$ mentioned in ~\cite{fionda2018community}, i.e., both of them have a tendency to spread more nodes of the target community into more communities. The main difference is that by using normalization, in our method, the number of new communities will not be too large, making the attack more concealed.


Finally, for target node attack, it is quite easy to hide a node by rewiring links. In this study, however, we try to hide a node by only adding links to the target node. We believe that adding, rather than deleting, certain links are much more convenient for social network users. For target node $t$, the fitness of chromosome $O_i$ thus is calculated by
\begin{eqnarray}
\phi_{i} =  \left\{
\begin{array}{cc}
\frac{d_t}{d_t+\beta|{O_i}} & \mbox{successful attack} \\
0 & \mbox{otherwise}
\end{array},\right.
\label{ti-1}
\end{eqnarray}
where $d_t$ is the degree of target node $t$ before the attack and $\beta|{O_i}$ is the length of chromosome $O_i$. Thus, $d_t+\beta|{O_i}$ is the degree of the target node after the attack represented by $O_i$.

For any kind of attack, after evaluating each individual in GA, we further use roulette selection method to select offspring. If the fitness of each individual in the population is $\phi_i$ $(i=1,2,\cdots,M)$ and the size of the population is $M$, the selection probability of individual $O_i$ is calculated by
\begin{eqnarray}
P(O_i) = \frac{\phi_{i}}{\sum^{M}_{j=1}\phi_{j}}.
\label{sel-1}
\end{eqnarray}

\subsection{Crossover\label{Cro}}
Traditionally, the length of chromosomes is set the same and is fixed throughout the evolution in GA. Here, we prefer to use \emph{non-equal} crossover, where the length of chromosomes can change in the process of crossover, so that we can find the approximate optimal solution with the smallest budget $\beta$, which is described by the following steps.
\begin{enumerate}
\item Choose two chromosomes $O_i$ and $O_j$ and form two subsets $E_i$=$\{A_i^E, B_i^E\}\subseteq{O_i}$ and $E_j$=$\{A_j^E, B_j^E\}\subseteq{O_j}$ by removing the identical elements that are considered nonexchangeable, where $A_i^E$ (or $A_j^E$) and $B_i^E$ (or $B_j^E$) are the \emph{exchangeable} gene sets of addition and deletion, respectively, in chromosomes $O_i$ (or $O_j$).
\item Calculate the lengths of $A_i^E$, $B_i^E$, $A_j^E$ and $B_j^E$, and denote them by $l_{A}^i$, $l_{B}^i$, $l_{A}^j$ and $l_{B}^j$, respectively.
\item Generate two random integers $r_i$ and $r_j$ in the intervals $[1, min (l_{A}^i,l_{B}^i)]$ and $[1, min (l_{A}^j,l_{B}^j)]$, respectively.
\item Suppose the numbers of rewired links by chromosomes $O_i$ and $O_j$ are $\beta_i$ and $\beta_j$, respectively, and the threshold is $\theta$. If $\beta_i-r_i+r_j$ or $\beta_j+r_i-r_j$ $\notin [1,\theta]$, go to step 3); otherwise, go to step 5).
\item select $r_i$ addition and deletion genes, respectively, from $O_i$, and select $r_j$ addition and deletion genes, respectively, from $O_j$; exchange the genes selected from $O_i$ with those selected from $O_j$, to generate two new chromosomes $\bar{O_1}$ and $\bar{O_2}$.
\end{enumerate}
The whole process of crossover is shown in Fig.~\ref{fig:crossover}.

\subsection{Mutation\label{Mut}}
Here, we further utilize the structural information around pairwise nodes as heuristic information to accelerate the searching process of GA.

For link addition, two nodes are of less similarity if the network distance (or the shortest path length) between them is relatively long. Therefore, in order to make the attack more effective, it is better to add the links between pairwise nodes of longer distance. Suppose the distance between a pair of disconnected nodes $i$ and $j$ is $\lambda_{ij}$, then the probability that a link addition gene to create a link between these two nodes is generated by mutation is defined as
\begin{eqnarray}
P(a_k) = \frac{{\lambda_{ij}}}{\sum_{(i,j)\notin{E}}{\lambda_{ij}}},
\label{mu-1}
\end{eqnarray}
where $a_k$ is the index of the pair of nodes $(i,j)$.

For link deletion, the links inside communities always have lower betweenness than those across different communities. Therefore, in order to make the attack more effective, it is better to delete the links of lower betweenness. Given a link with its index $b_k$, its betweenness is calculated by
\begin{eqnarray}
C_B(b_k) = \sum_{s,t \in V}\frac{\sigma(s,t|b_k)}{\sigma(s,t)},
\label{mu-3}
\end{eqnarray}
where $V$ is the node set in the network, $\sigma(s,t)$ represents the total number of shortest paths between nodes $s$ and $t$, and $\sigma(s,t|b_k)$ represents the number of shortest paths through the link. Then, the probability that a link deletion gene to remove the link is generated by mutation is defined as
\begin{eqnarray}
P(b_k) = \frac{\frac{1}{C_B(b_k)}}{\sum_{k=1}^m\frac{1}{C_B(b_k)}},
\label{mu-2}
\end{eqnarray}
where $m$ is the total number of links in the network.


\section{Experiments}
\label{sec:Experiments}
In this part, we will perform the three kinds of attacks on several synthetic networks and real-world networks. For global attack, we first compare our EPA with three heuristic algorithms under different budges $\beta$, and we also design the experiment to show the ability of EPA to find the approximate optimal budget. For target community attack, we first use EPA to get the approximate optimal budget and then compare EPA with the other algorithms under this budget. For target node attack, we select some representative nodes as targets. For each experiment, we run 10 times and record the mean result. Our experimental environment consists of i7-8700 3.2GHz (CPU), GTX 1050Ti 4GB (GPU), 16GB memory and Windows 10.


\subsection{Datasets}
To evaluate the attack effect of EPA, we use three community detection algorithms, i.e., greedy (GRE)~\cite{newman2004fast}, Infomap Algorithm (INF)~\cite{rosvall2008maps} and Louvain (LOU)~\cite{blondel2008fast}, on the six synthetic networks and three real networks, described as follows, with their basic properties presented in TABLE~\ref{table:sy_dataset} and TABLE~\ref{table:real_dataset}, respectively. The descriptions of parameters for generating a synthetic network are listed in TABLE~\ref{table:LFR} for convenience.
\begin{itemize}
\item \textbf{The synthetic networks~\cite{lancichinetti2008benchmark}:} These networks are generated by LFR benchmark, all of which are undirected and unweighted networks. 
\item \textbf{The USA college football (Football)~\cite{girvan2002community}:} This network represents the matches between American football teams during the season of 2000.
\item \textbf{email-Eu-core network (Email)~\cite{leskovec2007graph}:} The network represents the emails between members of a large European research institution.
\item \textbf{Political blogs (Pol.Blogs)~\cite{adamic2005political}:} This network represents the political leaning collected from blog directories.

\end{itemize}

\begin{table}[!ht]
\centering
\caption{The basic properties of the six synthetic networks.}
\begin{tabular}{@{}ccccccc@{}}
\toprule
     Network  & N   & k & $\max{k}$ & $\mu$ & $\min{c}$ & $\max{c}$   \\
     \midrule
  N-1000-$\mu$-0.3  &1000& 10 & 50 & 0.3 & 50 & 100 \\
  N-1000-$\mu$-0.5  &1000& 10 & 50 & 0.5 & 50 & 100 \\
  N-3000-$\mu$-0.3  &3000& 10 & 50 & 0.3 & 100 & 200 \\
  N-3000-$\mu$-0.5  &3000& 10 & 50 & 0.5 & 100 & 200 \\
  N-5000-$\mu$-0.3  &5000& 15 & 100 & 0.3 & 100 & 200 \\
  N-5000-$\mu$-0.5  &5000& 15 & 100 & 0.5 & 100 & 200 \\
  \bottomrule
\end{tabular}
\label{table:sy_dataset}
\end{table}

\begin{table}[!ht]
\centering
\caption{The basic properties of the three networks.}
\begin{tabular}{@{}ccccc@{}}
\toprule
     Network  & \#Nodes   & \#Links & \#Communities  \\
     \midrule
  Football  &115& 613 & 12 \\
  Email  &1005& 25571 & 42 \\
  Pol.Books &1490& 19090 & 2 \\
  \bottomrule
\end{tabular}
\label{table:real_dataset}
\end{table}

\begin{table}[!ht]
\centering
\caption{The meaning of parameters in synthetic networks.}
\begin{tabular}{@{}cc@{}}
\toprule
     Parameter  & Meaning  \\
     \midrule
  N  & number of nodes \\
  k  & average degree \\
  $\max{k}$  & maximum degree \\
  $\mu$  & mixing parameter \\
  $\min{c}$ & minimum community size \\
  $\max{c}$ & maximum community size \\
  \bottomrule
\end{tabular}
\label{table:LFR}
\end{table}

In experiments, we only consider undirected networks and remove the isolated nodes from data sets since they are meaningless in community detection.
The three community detection algorithms we adopt are also briefly introduced as follows to make the paper self-contained.
\begin{itemize}
\item \textbf{GRE~\cite{newman2004fast}:} Each node is considered as a separate community initially and the communities are fused in the direction of maximum increment of modularity $Q$. The complexity is $\mathcal{O}(|V|log^2(|V|))$.
\item \textbf{INF~\cite{rosvall2008maps}:} It's an information theory based algorithm which strive to compress the average description length for a random walk. Its complexity is $\mathcal{O}(|E|)$.
\item \textbf{LOU~\cite{blondel2008fast}:} This is a modularity based algorithm which can generate a hierarchical community structure by compressing the communities continuously. This algorithm runs in time $\mathcal{O}(|V|log(|V|))$.
\item \textbf{WAL~\cite{pons2005computing}:}  WalkTrap algorithm merges nodes based on the transition probability of random walk model. The complexity is $\mathcal{O}(|V|^2log(|V|))$.
\item \textbf{EIG~\cite{newman2006finding}:} This algorithm finds communities based on the fidler vector of Laplacian matrix. The complexity is $\mathcal{O}(|V|(|V|+|E|))$.
\item \textbf{SPI~\cite{reichardt2006statistical}:} it's a semi-supervised community detection algorithm which reduces community detection to the problem of finding the ground state of an infinite spin glass. Its complexity is $\mathcal{O}(|V|^{3.2})$.
\end{itemize}

\subsection{Baseline Attack Methods}
Inspired by various community detection algorithms, we use the following four attack methods as the baselines for global attack. Due to the poor performance of random attack, we introduce heuristic strategies on the random attack and propose two heuristic attack methods $A_B$ and $A_D$.
\begin{itemize}
\item \textbf{$A_Q$:} A GA based method where the modularity $Q$ is used to design the fitness function~\cite{chen2019ga}.
\item \textbf{$A_S$:} Rather than using the entropy-based fitness function, here we use the average of deception score~\cite{fionda2018community} as the fitness function and the rest is the same as EPA.
\item \textbf{$A_B$:} Deleting the links with the highest betweenness centrality, while adding the links between the nodes with the longest distance.
\item \textbf{$A_D$:} Deleting the links with the largest sum of degrees of their terminal nodes, while adding the links between the nodes with the longest distance.
\end{itemize}
For target community attack, we use the safeness based deception algorithm $D_{s}$ ~\cite{fionda2018community} and random algorithm $D_{w}$~\cite{waniek2018hiding} as the baseline attack methods, which can effectively hide the target community against different community detection algorithms, and they are briefly introduced as follows.
\begin{itemize}
\item \textbf{$D_{s}$:} Both link addition and deletion aim to maximize the safeness of target community defined in ~\cite{fionda2018community}. 
\item \textbf{$D_{w}$:} This method randomly adds links inter different communities while deletes links intra communities.
\end{itemize}
For target node attack, we propose a heuristic algorithm $D_{r}$ which randomly add links between target node and the nodes in other communities.

\subsection{Performance Metrics}
In order to verify the effectiveness of our EPA, we compare it with other baseline attack methods by using a series of metrics.

For global attack, we use Normalized Mutual Information (NMI) and Adjusted Rand Index (ARI) to evaluate the community detection results. Then, we further evaluate the attack effects by comparing their values before and after attacks. NMI and ARI are two of the most widely used metrics of community detection~\cite{liu2019evaluation,ghorbanian2015genetic}, and NMI is also used in community detection attack~\cite{fionda2018community,chen2019ga}. In particular, NMI and ARI are defined as
\begin{eqnarray}
\mbox{NMI}=\frac{-2\sum^{N_a}_{i=1}\sum^{N_b}_{j=1}m_{ij}\log(\frac{m_{ij}n}{M_{i}M_{j}})}{\sum^{N_a}_{i=1}M_i\log(\frac{M_i}{n})+\sum^{N_b}_{j=1}M_{j}\log(\frac{M_j}{n})},
\label{me-NMI}
\end{eqnarray}
\begin{eqnarray}
\mbox{ARI}=\frac{\sum_{ij}{m_{ij} \choose 2}-[\sum_i{M_{i} \choose 2}\sum_i{M_{j} \choose 2}]/{n \choose 2}}{\frac{1}{2}[\sum_i{M_{i} \choose 2}+\sum_j{M_{j} \choose 2}]-[\sum_i{M_{i} \choose 2}\sum_j{M_{j} \choose 2}]/{n \choose 2}},
\label{me-ARI}
\end{eqnarray}
where $M_{i}$ and $M_{j}$ are the sums over row $i$ and column $j$, respectively. $N_a$ and $N_b$ are the numbers of communities in control group (i.e., real community number) and test group (i.e., community number detected by certain community detection algorithm), respectively.

\begin{figure*}[!t]
\includegraphics[width=\linewidth]{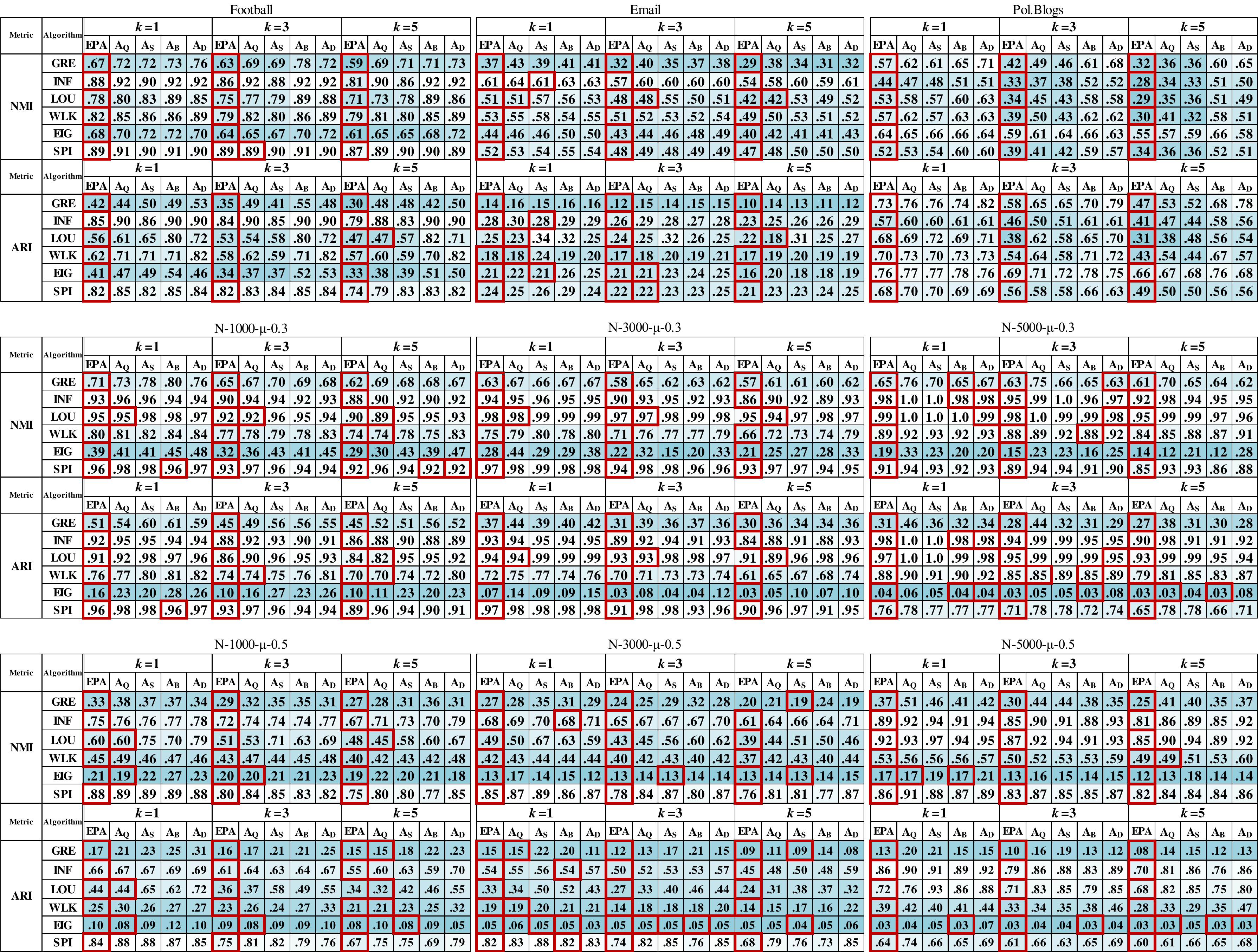}
\centering
\caption{The attack effects, reflected by the change of NMI and ARI, obtained by EPA and the four baseline attack methods by rewiring $k\%$ of links, with $k$ varies from 1 to 5, for the three real-world networks and six synthetic networks. Darker colors represent better attack performances and the best performance of each $k$ is shown in red box.}
\label{fig:value}
\end{figure*}

For target community attack, in addition to the fitness, we also use the deception score $H$ ~\cite{fionda2018community} to evaluate the attack effect, which is defined as
\begin{eqnarray}
H = \left[1-\frac{ \left| S(C) \right|-1}{\left| C \right|-1}\right] \times \left[\frac{1}{2}(1-\max(R))+\frac{1}{2}(1-\bar{P})\right],
\label{me-H}
\end{eqnarray}
where $\left| S(C) \right|$ is the number of connected components in the subgraph induced by the members in $C$. $\bar{P}$ and $R$ are the mean precision and recall rate, respectively.

For target node attack, we just use the percentage of target node degree increment in the attack to evaluate the results.

\begin{table*}[!t]
	\centering
	\caption{The community detection results before attack.}

	\begin{tabular}{@{}cccccccccccccccc@{}}
    \toprule
		\multirow{3}{*}{Network} & \multicolumn{2}{c}{GRE}    & \multicolumn{2}{c}{INF}    & \multicolumn{2}{c}{LOU} & \multicolumn{2}{c}{WAL}& \multicolumn{2}{c}{EIG}& \multicolumn{2}{c}{SPI} \\\cmidrule{2-13}
		&NMI&ARI &NMI&ARI &NMI&ARI &NMI&ARI &NMI&ARI &NMI&ARI\\
    \midrule
    Football&0.73&0.49&0.92&0.90&0.73&0.49 &0.89&0.82&0.70&0.46&0.91&0.85\\
    Email   &0.45&0.16&0.68&0.30&0.58&0.33 &0.57&0.19&0.50&0.26&0.55&0.25\\
    Pol.Blog&0.69&0.79&0.52&0.65&0.64&0.77 &0.69&0.78&0.64&0.76&0.63&0.77\\
    N-1000-$\mu$-0.3&0.73&0.54&0.97&0.95&0.99&0.98 &0.85&0.82&0.45&0.26&0.99&0.99\\
    N-1000-$\mu$-0.5&0.37&0.19&0.78&0.70&0.85&0.74 &0.47&0.27&0.23&0.10&0.92&0.90\\
    N-3000-$\mu$-0.3&0.63&0.34&0.96&0.96&1.0&1.0 &0.81&0.77&0.28&0.06&0.99&0.99\\
    N-3000-$\mu$-0.5&0.28&0.14&0.71&0.56&0.57&0.41 &0.44&0.20&0.14&0.05&0.89&0.85\\
    N-5000-$\mu$-0.3&0.75&0.44&1.00&1.00&1.0&1.0 &0.94&0.93&0.22&0.04&0.94&0.78\\
    N-5000-$\mu$-0.5&0.42&0.15&0.95&0.93&0.98&0.95 &0.56&0.44&0.21&0.04&0.89&0.66\\
      \bottomrule
	\end{tabular}
\label{table:bef_attack}
\end{table*}
\begin{figure*}[!t]
\includegraphics[width=\linewidth]{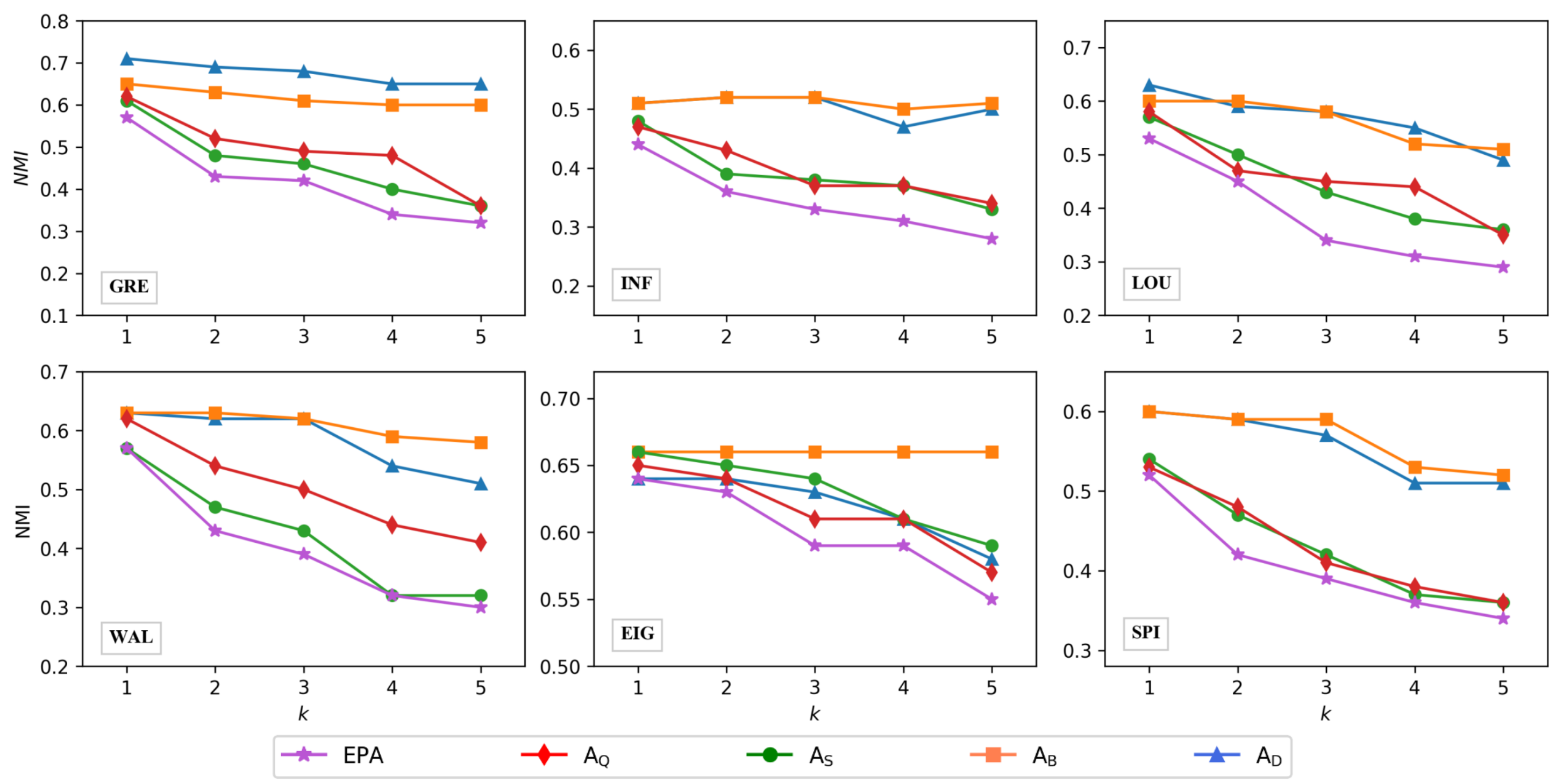}
\centering
\caption{ The attack effects of the five attack strategies on six community detection algorithms and Pol.Blogs network, in term of NMI, for various numbers of attacks.}
\label{fig:re_plot}
\end{figure*}
\subsection{Experimental Results\label{ER}}
\subsubsection{Global Attack}
In global attack, we fix the budget and rewire $k\%$ of links to compare our EPA with four baseline methods including $A_Q$, $A_S$, $A_B$ and $A_D$. We choose LOU as the basic community detection algorithm, and meanwhile we also use the GRE and INF algorithms to verify the black-box attack effect. The population used in the experiment is 100, the maximum number of iterations is 200, the crossover and mutation rates are 0.6 and 0.1, respectively. TABLE~\ref{table:bef_attack} represents the community detection results before attack.


The community detection results, in terms of NMI and ARI, on different datasets obtained by various community detection algorithms, after the attacks by EPA and the four baseline attack methods for various percentages of rewired links, are presented in Fig.~\ref{fig:value}. And the decrease of NMI on Pol.Blogs network is shown in Fig.~\ref{fig:re_plot}. Generally, we can find that, our EPA has the best attack effect on each performance metric, i.e., leading to smaller NMI and ARI, in most cases.

More specifically, we find that EPA performs especially well on Pol.Blogs network, i.e., rewiring $4\%$ links can decrease both NMI and ARI to near 0.3. This may be because LOU performs quite well in revealing the community structure of this network, and meanwhile this network is also easy to be disturbed since the connections are much sparse. In fact, none of the attack methods performs well on LFR generated networks with small mixing parameter, i.e., NMI and ARI still keep relatively high after any kind of attack, since the communities of these networks are relatively dense and thus are easier to be detected. Indeed, for larger mixing parameter, all attacks behave better, leading to smaller NMI and ARI. By comparison, EPA outperforms all the others on synthetic networks, no matter for large or small mixing parameters. Besides, we can also find that white-box attack (LOU) behaves better than black-box attacks (other community detection algorithms), which is quite intuitive since all the attack methods here are based on the LOU algorithm. More interestingly, by comparison, the communities detected by GRE are more affected than other algorithms, under the attack of LOU based EPA. This may be because both GRE and LOU are modularity-based algorithms, while the others are not. In addition, the communities detected by EIG on LFR networks are quite robust, which may be because EIG performs very poor compared with other community detection algorithms on LFR network.

Now, let's focus on the influence of the attenuation factor $c$ which controls the degree that the budget $\beta$ penalizes the fitness and ultimately controls the approximate optimal budget generate by EPA. We thus compare the attack results under different values of $c$, and fix the other parameters. For each value, we run the experiment 10 times and report the mean results in TABLE~\ref{table:re_c}, where we can see that as $c$ increases from 3 to 6, the final budget generated by EPA significantly decreases with a little bit sacrifice of attack effects, i.e., NMI and ARI increases slightly as $c$ increases. We thus suggest to use relatively large value of $c$ if we want to get a more concealed attack, while use relatively small value of $c$ if the approximate optimal attack effect is pursued.

\begin{table*}[!t]
	\centering
	\caption{The global attack results obtained by EPA under various attenuation factor $c$.}
	\begin{tabular}{@{}ccccccccccccc@{}}
    \toprule
		\multirow{2}{*}{Network} & \multicolumn{3}{c}{c=3}    & \multicolumn{3}{c}{c=4}    & \multicolumn{3}{c}{c=5} & \multicolumn{3}{c}{c=6} \\\cmidrule{2-13}
		&$\beta$&NMI&ARI &$\beta$&NMI&ARI &$\beta$&NMI&ARI &$\beta$&NMI&ARI\\
    \midrule
    Football&12.8&0.76&0.53&10.6&0.77&0.54&8.6&0.77&0.55&7.8&0.78&0.57 \\
    Email&566.6&0.48&0.22&506.4&0.49&0.23&134.5&0.51&0.25&113.7&0.53&0.26 \\
    Pol.Blogs&1947.1&0.33&0.35&1690.9&0.34&0.35&1172.8&0.37&0.36&1125.5&0.38&0.39 \\
      \bottomrule
	\end{tabular}
\label{table:re_c}
\end{table*}

\begin{figure}[!t]
\includegraphics[width=\linewidth]{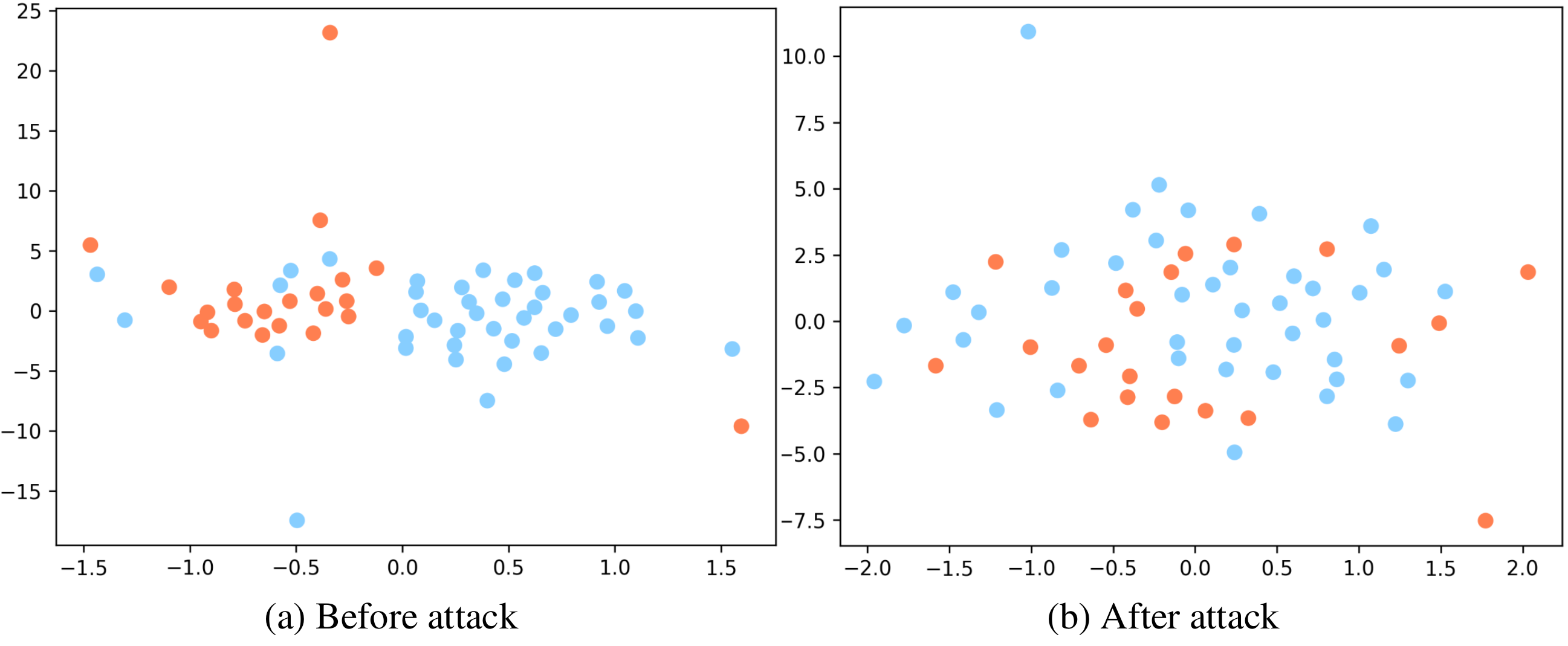}
\centering
\caption{ The t-SNE visualization of community deception obtained by EPA on Dolphin network. The nodes of same color belongs to the same community detected by GRE, and the target community is marked by blue. (a) Before attack; (b) After attack.}
\label{fig:TSNE}
\end{figure}

\subsubsection{Target Community Attack\label{TCA}}

Target community attack is also known as \emph{community deception}. As an example, we visualize the attack effect of our EPA on the Dolphin network using t-SNE algorithm~\cite{maaten2008visualizing}, as shown in Fig.~\ref{fig:TSNE}. Because the low-dimensional vectors generated by t-SNE algorithm will preserve the feature of network community structure, the position of nodes in Fig.~\ref{fig:TSNE} can reflect the result of community detection. Here, we compare EPA with safeness based deception algorithm $D_{s}$ ~\cite{fionda2018community} and random algorithm $D_{w}$ ~\cite{waniek2018hiding}. In order to make fair comparison, we fix the budget $\beta$ for each algorithm, i.e., we first use our EPA to find the approximate optimal  budget and then use it as the input of $D_{s}$ and $D_{w}$. Again, we record the mean values of fitness and deception score $H$ for all detected communities with at least 10 nodes on each network, as presented in TABLE~\ref{table:CH_lfr}-TABLE~\ref{table:CH_real}. Note that here, for synthetic networks, we only give the results of INF and LOU algorithms, since other algorithms is of high time complexity and thus is quite time-consuming on networks including more than thousands of nodes, especially we need to attack each community in a network.

\begin{table}[!t]
	\centering
	\caption{Deception results on synthetic networks.}
	\begin{tabular}{@{}cccccc@{}}
    \toprule
		\multirow{2}{*}{dataset} & \multirow{2}{*}{Alg}     & \multicolumn{2}{c}{INF }    & \multicolumn{2}{c}{Lou}
\\\cmidrule{3-6}
		&&Fitness&$H$&Fitness&$H$ \\
\midrule
    \multirow{3}{*}{1000-0.3}&EPA&\textbf{0.26}&\textbf{0.51}&\textbf{0.34}&\textbf{0.68} \\
                           &$D_{s}$&0.14&0.42&0.14&0.60\\
                            &$D_{w}$&0.04&0.18&0.08&0.55\\
\midrule
    \multirow{3}{*}{1000-0.5}&EPA	&\textbf{0.32}&\textbf{0.52}&\textbf{1.06}&\textbf{0.80} \\
    &$D_{s}$&0.12&0.36&0.54&0.61\\
    &$D_{w}$&0.10&0.32&0.52&0.61\\
    \midrule
    \multirow{3}{*}{3000-0.3}&EPA	&\textbf{0.16}&\textbf{0.41}&\textbf{0.26}&\textbf{0.72} \\
    &$D_{s}$&0.10&0.39&0.10&0.68\\
    &$D_{w}$&0.04&0.23&0.06&0.63\\
    \midrule
    \multirow{3}{*}{3000-0.5}&EPA	&\textbf{0.18}&\textbf{0.48}&\textbf{1.08}&\textbf{0.84} \\
    &$D_{s}$&0.12&0.40&0.50&0.60\\
    &$D_{w}$&0.10&0.39&0.50&0.60\\
    \midrule
    \multirow{3}{*}{5000-0.3}&EPA	&\textbf{0.06}&\textbf{0.27}&\textbf{0.08}&\textbf{0.51} \\
    &$D_{s}$&0.04&0.24&\textbf{0.08}&0.48\\
    &$D_{w}$&0.02&0.18&0.02&0.46\\
    \midrule
    \multirow{3}{*}{5000-0.5}&EPA	&\textbf{0.14}&\textbf{0.43}&\textbf{0.34}&\textbf{0.82} \\
    &$D_{s}$&0.10&0.37&0.12&0.70\\
    &$D_{w}$&0.04&0.26&0.08&0.67\\
      \bottomrule
	\end{tabular}
\label{table:CH_lfr}
\end{table}

\begin{table*}[!t]
	\centering
	\caption{Deception results on real-world networks.}
	\begin{tabular}{@{}cccccccccccccc@{}}
    \toprule
		\multirow{2}{*}{Network} & \multirow{2}{*}{Alg} & \multicolumn{2}{c}{GRE}    & \multicolumn{2}{c}{INF }    & \multicolumn{2}{c}{Lou} & \multicolumn{2}{c}{WAL}& \multicolumn{2}{c}{EIG}& \multicolumn{2}{c}{SPI} \\\cmidrule{3-14}
		&&Fitness&$H$&Fitness&$H$&Fitness&$H$&Fitness&$H$&Fitness&$H$&Fitness&$H$ \\
\midrule
    \multirow{3}{*}{Football}&EPA	&\textbf{0.62}&\textbf{0.58}&\textbf{0.48}&\textbf{0.41}&\textbf{0.54}&\textbf{0.46} &\textbf{0.33}&\textbf{0.41}&\textbf{0.66}&\textbf{0.56}&\textbf{0.29}&\textbf{0.34}\\
    &$D_{s}$&0.26&0.33&0.04&0.08&0.10&0.16&0.18&0.32&0.38&0.40&0.15&0.27\\
    &$D_{w}$&0.16&0.22&0.00&0.00&0.06&0.09&0.05&0.08&0.30&0.34&0.05&0.09\\
\midrule
    \multirow{3}{*}{Email} &EPA	&\textbf{0.68}&\textbf{0.57}&\textbf{0.54}&\textbf{0.45}&\textbf{0.78}&\textbf{0.66} &\textbf{0.48}&\textbf{0.44}&\textbf{0.68}&\textbf{0.55}&\textbf{0.34}&\textbf{0.49}\\
    &$D_{s}$&0.40&0.52&0.06&0.13&0.46&0.54&0.28&0.34&0.34&0.49&0.28&0.41\\
    &$D_{w}$&0.32&0.44&0.02&0.06&0.36&0.47 &0.13&0.25&0.30&0.41&0.27&0.43\\
    \midrule
    \multirow{3}{*}{Pol.Blogs} &EPA	&\textbf{0.38}&\textbf{0.47}&\textbf{0.76}&\textbf{0.58}&\textbf{0.54}&\textbf{0.53}
    &\textbf{0.56}&\textbf{0.39}&\textbf{0.93}&\textbf{0.42}&\textbf{0.42}&\textbf{0.44} \\
    &$D_{s}$&0.26&0.43&0.18&0.52&0.32&0.42 &0.51&0.32&0.67&0.33&0.32&0.38\\
    &$D_{w}$&0.22&0.40&0.10&0.33&0.30&0.38&0.49&0.24&0.43&0.29&0.33&0.38\\
      \bottomrule
	\end{tabular}
\label{table:CH_real}
\end{table*}

We find that, again, our EPA behaves significantly better, in terms of much higher fitness and deception score $H$, than both $D_{s}$ and $D_{w}$. And such results are quite robust. For instance, for the Football network, it seems that $D_{s}$ and $D_{w}$ lose their attack effects on INF and LOU, i.e., the values of Fitness and $H$ are quite small by comparing with those on GRE, as presented in TABLE~\ref{table:CH_real}. However, by using our EPA, their values still keep relatively large for all the six community detection algorithms, indicating that EPA is effective in both white-box and black-box situations. In addition, we can also find that EPA is not effective on SPI, which may be because SPI is a semi-supervised algorithm with relatively strong robustness. Moreover, it seems that the synthetic networks with smaller mixing parameters $\mu$ always have lower fitness in the corresponding cases. The reason may be that the networks of smaller $\mu$ tend to greater differences between communities, making it more difficult to hide the target community, i.e., corresponding to a smaller fitness value, according to Eq.~(\ref{tc-1}).

Compared with $D_{s}$ algorithm, EPA algorithm needs complete knowledge about network structure, but the effect of community deception is much better than $D_{s}$ algorithm. In addition, because the fitness of EPA for target community attack only considers the change of the target community, it can attack the target community as long as the community detection algorithm can accurately divide the target community before attack. In the other word, EPA can still attack the target community even if the knowledge of other communities is missing partly.



\subsubsection{Target Node Attack\label{TNA}}
\begin{figure}[!t]
\includegraphics[width=\linewidth]{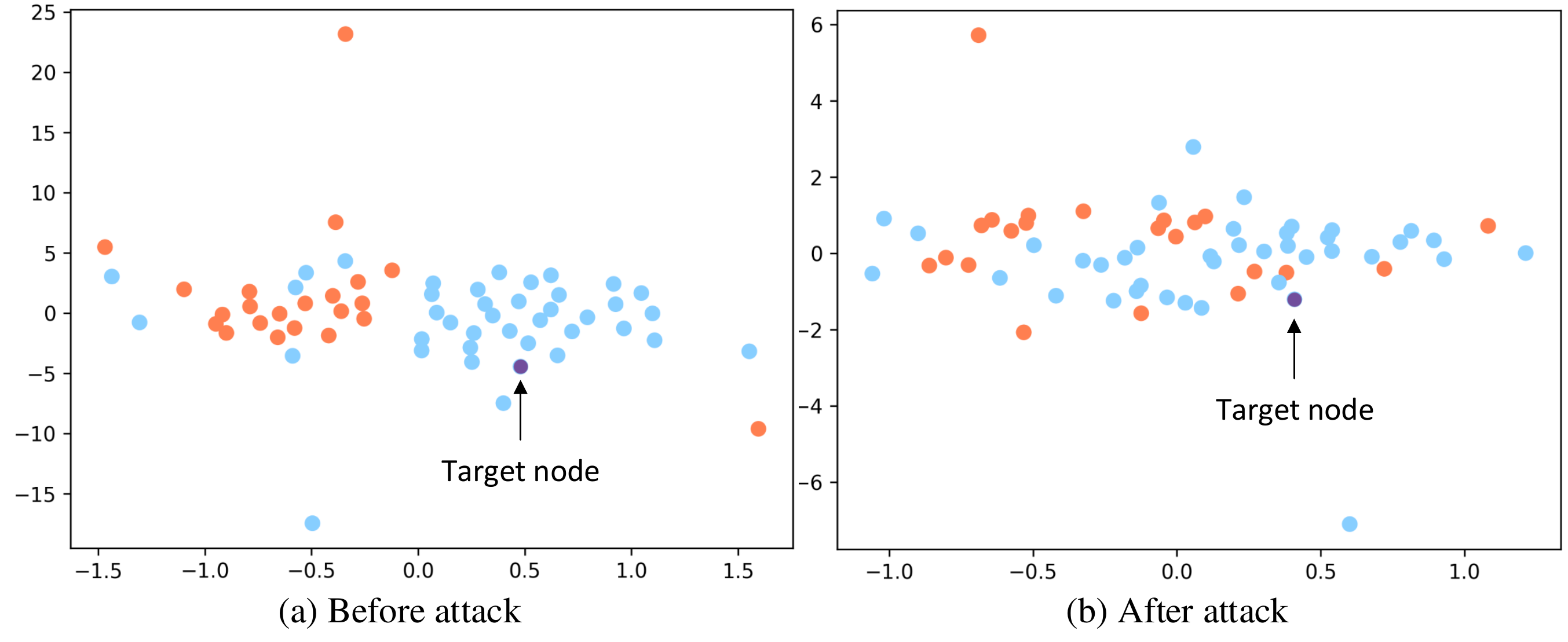}
\centering
\caption{The t-SNE visualization of target node attack obtained by EPA on Dolphin network. The target node originally belongs to the blue community. After the attack, it belongs to the red community. The communities are detected by GRE. (a) Before attack; (b) After attack.}
\label{fig:TSNEIn}
\end{figure}

Similarly, for target node attack, we also visualize the attack effect of our EPA on the Dolphin network using t-SNE algorithm, as shown in Fig.~\ref{fig:TSNEIn}. In addition to the three real-wrold datasets, we also do target node attack on the 9/11 terrorist networks~\cite{krebs2002mapping}, in which terrorists may try to hide themselves in terrorist community. The 9/11 terrorist network consists of 37 nodes and 85 links. Here, we focus on attacking hub and bridge nodes in a network since they are always more important in various applications. In particular, we first rank the nodes in each network based on their degree and betweenness, from large to small, respectively. Then, in addition to selecting the two nodes with the largest degree and betweenness, respectively, as the target nodes, we also sum the two orders for each node and choose the top one as our another target node. We name these three
target nodes as $T_1$,$T_2$ and $T_3$, respectively. Finally, we compare EPA with the $D_r$ algorithm that randomly adds links between target nodes and the nodes in different communities.


\begin{table}[!t]
	\centering
	\caption{Target node attack on synthetic networks.}
	\begin{tabular}{@{}cccccc@{}}
    \toprule
		\multirow{2}{*}{dataset} & \multirow{2}{*}{Node}     & \multicolumn{2}{c}{INF }    & \multicolumn{2}{c}{Lou}
\\\cmidrule{3-6}
		&&EPA&$D_r$&EPA&$D_r$ \\
\midrule
    \multirow{3}{*}{1000-0.3}&$T_1$&\textbf{67\%}&336\%&\textbf{53\%}&406\% \\
                           &$T_2$&\textbf{64\%}&412\%&\textbf{43\%}&349\%\\
                            &$T_3$&\textbf{70\%}&409\%&\textbf{55\%}&445\%\\
\midrule
    \multirow{3}{*}{1000-0.5}&$T_1$	&\textbf{43\%}&385\%&\textbf{53\%}&227\% \\
    &$T_2$&\textbf{46\%}&320\%&\textbf{22\%}&58\%\\
    &$T_3$&\textbf{29\%}&265\%&\textbf{28\%}&117\%\\
    \midrule
    \multirow{3}{*}{3000-0.3}&$T_1$	&\textbf{142\%}&340\%&\textbf{40\%}&625\% \\
    &$T_2$&\textbf{114\%}&408\%&\textbf{47\%}&522\%\\
    &$T_3$&\textbf{194\%}&493\%&\textbf{54\%}&596\%\\
    \midrule
    \multirow{3}{*}{3000-0.5}&$T_1$	&\textbf{242\%}&302\%&\textbf{16\%}&138\% \\
    &$T_2$&\textbf{60\%}&279\%&\textbf{8\%}&19\%\\
    &$T_3$&\textbf{78\%}&321\%&\textbf{16\%}&26\%\\
    \midrule
    \multirow{3}{*}{5000-0.3}&$T_1$	&\textbf{126\%}&625\%&\textbf{264\%}&1853\% \\
    &$T_2$&\textbf{97\%}&535\%&\textbf{224\%}&1280\%\\
    &$T_3$&\textbf{338\%}&691\%&\textbf{416\%}&2261\%\\
    \midrule
    \multirow{3}{*}{5000-0.5}&$T_1$	&\textbf{60\%}&560\%&\textbf{55\%}&706\% \\
    &$T_2$&\textbf{70\%}&558\%&\textbf{52\%}&653\%\\
    &$T_3$&\textbf{128\%}&672\%&\textbf{83\%}&815\%\\
      \bottomrule
	\end{tabular}
\label{table:IH_lfr}
\end{table}

\begin{table*}[!t]
\centering
\caption{Target node attack on real-world networks.}
\begin{tabular}{@{}cccccccccccccc@{}}
\toprule
\multirow{2}{*}{Network} &\multirow{2}{*}{Node} & \multicolumn{2}{c}{GRE}& \multicolumn{2}{c}{INF}    & \multicolumn{2}{c}{LOU} & \multicolumn{2}{c}{WAL}& \multicolumn{2}{c}{EIG}& \multicolumn{2}{c}{SPI} \\\cmidrule{3-14}
		&&EPA&$D_r$&EPA&$D_r$&EPA&$D_r$& EPA& $D_r$&EPA&$D_r$ &EPA&$D_r$\\
     \midrule
    \multirow{3}{*}{Football}&$T_1$&\textbf{8.3\%}&186\%&\textbf{67\%}&367\%&\textbf{33\%}&658\% &\textbf{42\%}&67\%&\textbf{8.3\%}&17\%&\textbf{67\%}&375\% \\
    &$T_2$&\textbf{9.1\%}&155\%&\textbf{27\%}&182\%&\textbf{18\%}&145\% &\textbf{9.1\%}&45\%&\textbf{9.1\%}&18\%&\textbf{17\%}&47\%\\
    &$T_3$&\textbf{33\%}&199\%&\textbf{50\%}&550\%&\textbf{42\%}&237\% &\textbf{25\%}&50\%&\textbf{17\%}&33\%&\textbf{38\%}&52\%\\
        \midrule
    \multirow{3}{*}{Email}&$T_1$&\textbf{8.4\%}&15\%&\textbf{10\%}&35\%&\textbf{0.3\%}&0.87\% &\textbf{19\%}&122\%&\textbf{4.3\%}&18\%&\textbf{1.7\%}&11\% \\
    &$T_2$&\textbf{117\%}&23\%&\textbf{6.0\%}&29\%&\textbf{0.5\%}&2.31\% &\textbf{17\%}&116\%&\textbf{1.4\%}&2.9\%&\textbf{1.4\%}&10\%\\
    &$T_3$&\textbf{123\%}&32\%&\textbf{23\%}&61\%&\textbf{5.2\%}&12\% &\textbf{25\%}&123\%&\textbf{24\%}&39\%&\textbf{16\%}&34\%\\
     \midrule
     \multirow{3}{*}{Pol.Blogs}&$T_1$&\textbf{39\%}&83\%&\textbf{34\%}&51\%&\textbf{37\%}&52\% &\textbf{16\%}&30\%&\textbf{109\%}&118\%&\textbf{20\%}&66\%\\
    &$T_2$&\textbf{38\%}&75\%&\textbf{16\%}&33\%&\textbf{17\%}&27\% &\textbf{38\%}&47\%&\textbf{7.6\%}&121\%&\textbf{32\%}&41\%\\
    &$T_3$&\textbf{32\%}&71\%&\textbf{24\%}&60\%&\textbf{26\%}&48\%  &\textbf{20\%}&25\%&\textbf{211\%}&227\%&\textbf{12\%}&61\%\\
    \midrule
    \multirow{3}{*}{TER}&$T_1$&\textbf{6.7\%}&55\%&\textbf{27\%}&93\%&\textbf{13\%}&39\% &\textbf{47\%}&72\%&\textbf{13\%}&94\%&\textbf{27\%}&69\%\\
    &$T_2$&\textbf{False}&False&\textbf{30\%}&84\%&\textbf{30\%}&42\% &\textbf{30\%}&71\%&\textbf{10\%}&182\%&\textbf{20\%}&32\%\\
    &$T_3$&\textbf{14\%}&41\%&\textbf{50\%}&121\%&\textbf{14\%}&44\%  &\textbf{29\%}&54\%&\textbf{14\%}&38\%&\textbf{21\%}&54\%\\
  \bottomrule
\end{tabular}
\label{table:IH_real}
\end{table*}

The percentages of degree increment in the attack for different strategies are presented in TABLE~\ref{table:IH_lfr}-TABLE~\ref{table:IH_lfr}. We can find that, in general, EPA performs significantly better than $D_r$, in terms of smaller percentage of degree increment, especially in the football network. Moreover, the bridge nodes of large betweenness are relatively easy to be attacked, since the neighbors of bridge nodes are always distributed in different communities, making them quite sensitive to link changes. On the contrary, it's relatively difficult to hide a node with both high degree and betweenness, since as hub nodes, a much large number of links always need to be added to change their communities. In addition, we can also find that both EPA and $D_r$ algorithms fail to attack GRE algorithm, which means that the community that target node $T_2$ belongs to can't be changed even if $T_2$ is connected to a number of nodes in other communities. It may be due to the fact that, based on GRE algorithm, the community that $T_2$ belongs to is not only very densely connected, but also all the other nodes except $T_2$ of this community has few links to the other communities. Note that, for real-world networks, based on EPA, the number of added links is much smaller than the degree of target nodes, indicating our method performs well in target node attack.

\section{Conclusions}
\label{sec:Conclusions}
In this paper, we propose a novel Evolutionary Perturbation Attack (EPA) method, based on Genetic Algorithm (GA), to disturb community detection algorithms in three scales, from local to global, by only changing a small fraction of links. In particular, we design a non-equal crossover strategy to treat the length of chromosome as a variable and naturally integrate it into GA; and we also integrate the network structural information, such as network distance and betweenness, into the algorithm to guide the search so as to find the better approximate optimal attack strategy more quickly. Numerical experiments on six synthetic networks and three real-world validate the effectiveness and transferability of our EPA method on attacking various community detection algorithms, i.e., by comparing with other attack methods of different scales, EPA behaves the best in most cases, achieving the state-of-the-art attack effects. However, our EPA is of relatively high time complexity, making it time consuming to deal with large-scale networks.

In the future, we will expand this work in the following three directions. First, we will propose new network coding methods and also utilize more network structural properties to improve the efficiency of EPA; second, we will try to integrate network embedding and deep learning graph models to improve the attack performance; third, we will do more experiments on more various networks, and further check the effectiveness of our EPA on the downstream algorithms based on community detection, to see the potential influence of EPA on many real applications.

\section*{Acknowledgment}
The authors would like to thank all the members in our IVSN research group in Zhejiang University of Technology for the valuable discussion about the ideas and technical details presented in
this paper.



\ifCLASSOPTIONcaptionsoff
  \newpage
\fi




\bibliographystyle{IEEEtran}
\bibliography{refnew}

\vfill




\end{document}